\documentclass[%
 reprint,
 amsmath,amssymb,
 aps,
 pre,
]{revtex4-1}

\usepackage{graphicx}
\usepackage{dcolumn}
\usepackage{bm}
\usepackage{url}

\usepackage{amsmath}
\usepackage{amssymb} 
\usepackage{mathtools}

\usepackage{appendix}
\usepackage[bf]{subfigure}
\usepackage[utf8]{inputenc}
\newcommand{\inst}[1]{$^#1$}
\usepackage{tikz}
\usepackage{indentfirst}
\usepackage{enumitem}  
\usepackage{placeins}
\usepackage{appendix}
\usepackage{lipsum}
 
\newcommand{\acoef}[0]{a_i}
\newcommand{\nnodes}[0]{V}
\newcommand{\nedges}[0]{M}
\newcommand{\npop}[0]{N_{pop}}
\newcommand{\ltcoef}[0]{l}  
\newcommand{\ithreshold}[0]{\varepsilon_I}
\newcommand{\herdlim}[0]{\rho_{herd}}
\newcommand{\peaksize}[0]{\rho_I^{\text{max}}}
\newcommand{\outbsize}[0]{\rho_R^{\text{max}}}
\newcommand{\aimpulse}[0]{J}
\newcommand{\abarmax}[0]{\overline{a}_{\text{max}}}
\newcommand{\outbthresin}[0]{\phi_{\text{in}}}
\newcommand{\outbthresout}[0]{\phi_{\text{out}}}

\begin{document}
\preprint{APS/123-QED}

\title{Modeling the effects of social distancing on the large-scale spreading of diseases}

\author{Paulo Cesar Ventura\inst{1}}
\email[Corresponding author: ]{paulo.ventura.silva@usp.br}
\altaffiliation[]{Av. Trabalhador Sancarlense, 400, 13566590, S\~{a}o Carlos - SP, Brazil }
\author{Alberto Aleta\inst{2}}
\author{Francisco A. Rodrigues\inst{3}}
\author{Yamir Moreno\inst{{2,4}}}

\affiliation{
  \inst{1} Instituto de F\'{i}sica de  S\~{a}o Carlos, Universidade de S\~{a}o Paulo, S\~{a}o Carlos, SP, Brazil.\\
  \inst{2} ISI Foundation, Turin, Italy\\
  \inst{3} Instituto de Ci\^{e}ncias Matem\'{a}ticas e de Computa\c{c}\~{a}o, Universidade de S\~{a}o Paulo, S\~{a}o Carlos, SP, Brazil.\\
  \inst{4}Institute for Biocomputation and Physics of Complex Systems (BIFI), 50018 Zaragoza, Spain\\
  \inst{5}Department of Theoretical Physics, University of Zaragoza, 50018 Zaragoza, Spain \\
}

\begin{abstract}
To contain the propagation of emerging diseases that are transmissible from human to human, non-pharmaceutical interventions (NPIs) aimed at reducing the interactions between humans are usually implemented. One example of the latter kind of measures is social distancing, which can be either policy-driven or can arise endogenously in the population as a consequence of the fear of infection. However, if NPIs are lifted before the population reaches herd immunity, further re-introductions of the pathogen would lead to secondary infections. Here we study the effects of different social distancing schemes on the large scale spreading of diseases. Specifically, we generalize metapopulation models to include social distancing mechanisms at the subpopulation level and model short- and long-term strategies that are fed with local or global information about the epidemics. We show that different model ingredients might lead to very diverse outcomes in different subpopulations. Our results suggest that there is not a unique answer to the question of whether contention measures are more efficient if implemented and managed locally or globally and that model outcomes depends on how the full complexity of human interactions is taken into account. 
\end{abstract}

\maketitle

\section{Introduction}

The spreading of infectious diseases is a complex process involving two main aspects. On the one hand, the spreading capabilities of the pathogen depend on its biological properties, the characteristics of the host, and the many environmental factors that can play an important role. On the other hand, the pathogen (except for vector-borne diseases) can only spread if two hosts have some kind of contact. As such, the behavior of the host is a key element in the study of epidemic spreading \cite{Funk2010Sep,Read2012Dec}. For this reason, many researches have focused on studying how animals and humans interact, in order to inform mathematical models and produce better forecasts \cite{Wilson2019,Mossong2008Mar,Arregui2018Dec,Salathe2010Dec}.

Despite having acknowledged the role that behavior plays in the spreading of infectious diseases, epidemic models usually neglect the possibility that hosts will change their behavior due to an ongoing outbreak \cite{eksin2019systematic}. An exception to this are awareness models, in which both an epidemic and information spread at the same time in the population and host reacts accordingly (see \cite{daSilva2019Sep} and the references therein). These behavioral changes have been observed both in animal \cite{Stroeymeyt2018Nov} and human societies \cite{SteelFisher2010Jun}. More recently, the spreading of COVID-19 has clearly demonstrated the variety of ways in which humans react to an epidemic. For instance, without forceful government intervention, traffic in Seoul's subway declined sharply following the first deaths in South Korea \cite{Park2020Apr}. Conversely, several communities in other regions of the world defied social distancing measures \cite{Waitzberg2020Dec,Gollwitzer2020Nov}.

Public health authorities base many decisions on the forecasts produced by epidemic models. It is thus of paramount importance to include the effect of behavioral changes as something inherently attached to the spreading of the epidemic \cite{eksin2019systematic}. Of particular interest in this regard are metapopulation models. These models provide a simple way of incorporating the spatial heterogeneity of human societies, while keeping them mathematically tractable. In essence, in a metapopulation model hosts are grouped in different subpopulations, and the exchange of individuals between subpopulations is governed by certain rules. Within each subpopulation it is possible to assume that the spreading from host to host follows the classical homogeneous mixing approach \cite{Lloyd1996Mar,Colizza2008Apr,Aleta2017Mar}, or include more details on individual heterogeneities through different techniques \cite{Ajelli2010Dec,Apolloni2014Dec}.

These models have been extensively used during the emergence of new pathogens to study how an outbreak might propagate globally \cite{Bajardi2011Jan}. In the particular case of COVID-19, they have also been used to study the early spreading of the disease across countries \cite{Chinazzi2020Apr}, but also within countries \cite{Costa2020Dec,Aleta2020Oct,Aleta2020Dec}. Yet, these models usually focus on the effect of varying the rules governing the flux of individuals between subpopulations, or on globally modifying the transmission due to the introduction of public health interventions. In this paper, our objective is to understand the role that social distancing can play in these models and shed some light on the effect that it can have on the spatial spreading of epidemics when the response is heterogeneous across regions.

\section{Model description}
\label{sec:description}

We implement a discrete and stochastic SIR-metapopulation model composed by $V$ subpopulations \cite{Keeling2004Jan,Ball2015Mar,Wang2014Oct}. In each subpopulation, individuals can interact with each other and spread the disease following the classical SIR model. These subpopulations are connected through a certain network, so that an individual may travel to another subpopulation only if it exists a direct link between the source and target subpopulations.

To include social distancing effect, either policy-driven or as a consequence of fear of infection, we consider an additional coefficient that modifies the transmissibility of individuals. This coefficient mimics the behavioral responses of the population, and evolves according to the densities of infectious and recovered individuals \cite{eksin2019systematic}. In section \ref{sec:analyt}, we study this model in an homogeneously mixed population. For the remaining sections, we apply the model to a metapopulation system.

\begin{table}[]
    \begin{ruledtabular}
    \centering
    \begin{tabular}{c|c}
        Description&Symbol\\
        \hline
        Individual infection probability    & $\beta$ \\
        Recovery probability/rate            & $\mu$   \\
        Basic reproduction number           & $R_0 = \beta / \mu$     \\
        Contact reduction coefficient       & $\acoef{}$ \\
        Response strength exponent          & $k$ \\
        Long-term coefficient               & $\ltcoef$  \\
        Mobility coefficient                & $\tau$     \\
        Threshold for infectious fraction   & $\ithreshold$ \\
        \\
        Individual counts in each state in subpopulation $i$  & $S_i, I_i, R_i$ \\
        Global individual counts in each state  & $N_S, N_I, N_R$ \\
        Global fractions in each state      & $\rho_S, \rho_I, \rho_R$ \\
        \\
        Number of nodes (subpopulations) on network & $\nnodes{}$ \\
        Number of links on network          & $\nedges$  \\
        Number of individuals in subpopulation $i$     & $N_i$     \\
        Total number of individuals in population & $\npop$ \\
        Maximum distance for connection on RGN & $d$  \\
        Link weight between subpopulations $i$ and $j$   & $T_{ij}$ \\
        \\
        Long-term strategy                  & LT \\
        Short-term strategy                 & ST \\
         network            & RGN \\ 
        
    \end{tabular}
    \caption{List of symbols and acronyms.}
    \label{tab:my_label}
\end{ruledtabular}
\end{table}

\subsection{Epidemic spreading}
\label{sec:description_epid}


\begin{figure*}[t]
    \centering
    \includegraphics[width=0.70\textwidth]{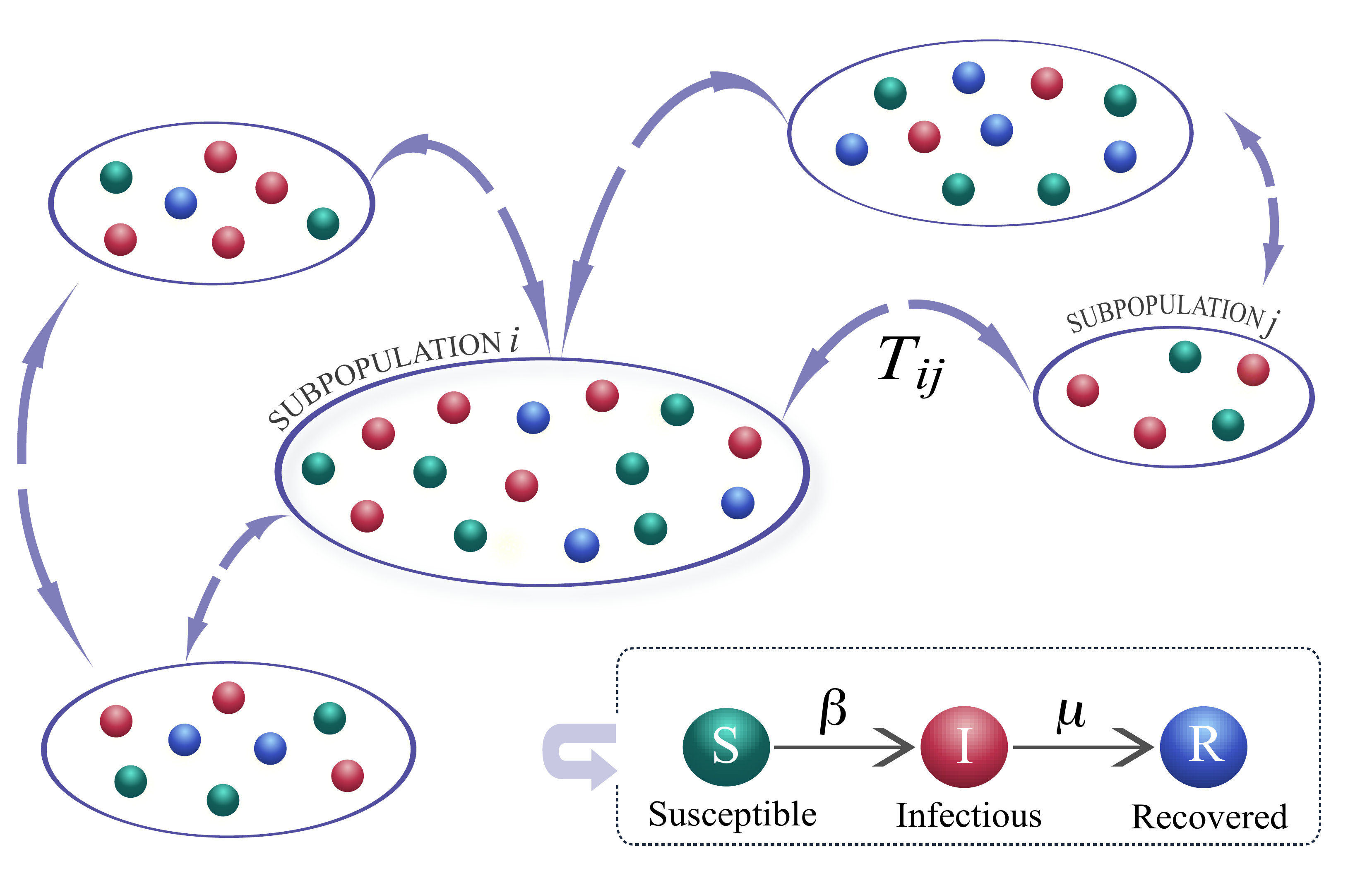}
    \caption{Scheme of the epidemic and mobility models in a metapopulation. The SIR epidemic dynamics occur inside each subpopulation, where homogeneous mixing is assumed. Also at each time step, individuals move between neighboring subpopulations $i$ and $j$ according to a mobility matrix $T_{ij}$.}
    \label{fig:model_scheme}
\end{figure*}

For the SIR compartmental model, individuals are assigned compartments according to their infectious status: susceptible (S) if they do not have the disease and can catch it; infectious (I) when they have the disease and can transmit it to susceptibles, and removed (R) when they no longer transmit the disease after being infectious (either by recovery or deceasing). At each model's epidemic update, the following rules determine the transitions between compartments:

\begin{itemize}
    
    \item \textbf{S $\rightarrow$ I}: a susceptible individual in subpopulation $i$ can become infectious with probability $P_i(S \rightarrow I) = 1 - (1 - a_i \beta / N_i)^{I_i}$, where $\beta$ is the individual transmission probability, $N_i$ is the number of individuals in subpopulation $i$ and $I_i$ is the number of infectious individuals in the same subpopulation. The term $\acoef$ is called the \emph{coefficient of contact reduction}, and depends on the considered strategy as we describe later on.
        
    
    \item \textbf{I $\rightarrow$ R}: an infectious individual is moved to the R (removed) compartment with probability $\mu$, which is the inverse of the average infectious period. After this event, the individual no longer participates on the epidemic dynamics.
\end{itemize}

Throughout this paper, we set $\mu$ to $1 / 4$ and the basic reproductive number $R_0 = \beta / \mu$ to $1.5$, which is compatible with the parameters of an influenza-like disease.
We define $N_S = \sum_{i=1}^{\nnodes}S_i$ as the total number of susceptible individuals in the whole population and, analogously, $N_I$ and $N_R$ for the infectious and removed compartments.
The coefficient of contact reduction $\acoef$ is what determines the behavioral responses to the epidemics, and we consider different scenarios for such a response, each one with a different definition of $\acoef$:

\begin{enumerate}
    \item \textbf{Social distancing based on global information}
    
    This scenario is based on the one proposed by Eksin \emph{et al.} in \cite{eksin2019systematic}, which here we extend to metapopulations. The coefficient of contact reduction emulates the social response to an increase in the number of infections, and it is a function of the total (global) number of infected and recovered individuals, given by:
    
    \begin{equation}
        \label{eq:global_acoef}
        \acoef = \left( 1 - \frac{(N_I + \ltcoef \cdot N_R)}{N} \right) ^ k
    \end{equation}
    
    Since $a_i$ only depends on global quantities, its value is the same for all subpopulations. The parameter $k$ is the \emph{response strength}, an adjustable exponent that calibrates the intensity of the response against the number of cases. The coefficient $\ltcoef$ is called \emph{coefficient of long-term}, and determines the importance of the R compartment to the behavioral response. Following \cite{eksin2019systematic}, $l = 0$ represents the \emph{short-term strategy} (ST), where the response weakens when the incidence drops. The case $l = 1$ is the \emph{long-term strategy} (LT), in which the awareness is proportional to the total prevalence. These two limiting cases differ in many aspects, which we describe throughout the rest of the paper.
    
    \item \textbf{Social distancing based on local information}
    
    In this variant of the previous scenario, we consider that each subpopulation (node) responds individually, according to its own number of cases. The expression for the coefficient of contact reduction is:
    
    \begin{equation}
        \label{eq:local_acoef}
        \acoef = \left(1 - \frac{(I_i + \ltcoef \cdot R_i)}{N_i}\right) ^ k
    \end{equation}

    \item \textbf{Constant response after threshold}
    
    For comparison, we also implement a more traditional scenario in which the transmissibility is reduced by a constant factor $a_0$ once the overall number of cases $N_I + N_R$ overcomes a given threshold $N \cdot \ithreshold{}$ (with $0 \leq \ithreshold{} < 1$). Here we use only the global information, thus $\acoef{}$ is defined as:
    
    \begin{equation}
         \acoef =\begin{cases}1 & N_I + N_R < N\ithreshold{} \\ a_0 & N_I + N_R \geq N\ithreshold{} \end{cases} 
    \end{equation}
    
    and is the same for all subpopulations. This scenario can be considered as a baseline in which social distancing (or governments policies) is constant and independent of the state of the system, 
    and it is regarded here as a benchmark that is often assumed as the behavioral response in metapopulation epidemics. \cite{Perra2011Aug,Aleta2020Dec,Calvetti2020Jun,Manfredi2013}.

\end{enumerate}

\subsection{Mobility model}

Following a standard metapopulation framework, the mobility between subpopulations is modeled as a random walk through the links of a graph of $\nnodes{}$ nodes and $\nedges{}$ links. It is controlled by a master parameter $\tau$ called the \emph{mobility coefficient}, as well as the weights $T_{ij}$ of existing links between subpopulations (nodes) $i$ and $j$. 
During a mobility update, each individual in subpopulation $i$ travels to subpopulation $j$ with probability $p_{ij} = \tau \cdot T_{ij} / N_i(0)$, where $N_i(0)$ is the number of individuals attributed to subpopulation $i$ at the initial time step. Within this scheme, we have the following features:

\begin{itemize}
    \item The average number of individuals expected to travel from $i$ to $j$ is $\tau \cdot T_{ij}$.
    
    \item If all links are reciprocal and symmetric (i.e., $T_{ji} = T_{ij}$ for all connected pairs of subpopulations $i, j$), then the net fluxes between the subpopulations are balanced and the population of each one remains approximately constant, fluctuating around $N_i(0)$. We use this configuration throughout the paper.
    
    \item The probability that an individual in subpopulation $i$ travels anywhere is $p_i = \tau \sum_j T_{ij} / N_i(0)$, and may vary for each subpopulation. We always set $\tau$ to be small enough so that $p_i < 1$ for every subpopulation $i$.

\end{itemize}

During a single time step, we first perform the epidemic interactions in each subpopulation, then update the number of individuals that are in each state. After this, we apply the mobility rules to determine how many individuals move through each link, and update the actual numbers only after all fluxes have been calculated. This way, the results do not depend on the order at which we ``visit'' the subpopulations to perform the calculations. The mobility and epidemic models are schematically represented in Figure \ref{fig:model_scheme}.


\section{Analytical insights for homogeneously mixed populations}
\label{sec:analyt}

For sufficiently low mobility between subpopulations, the local dynamics can be well described by an isolated homogeneously mixed system. Also for sufficiently high number of individuals, we can use rate equations for the expected fractions of the population in each compartment, substituting $S/N$, $I/N$ and $R/N$ by $\rho_S$, $\rho_I$ and $\rho_R$, respectively. The chosen expression for the behavioral response mechanism is simple enough to allow for analytical manipulation. Particularly, in the long-term (LT) strategy (i.e., with $l = 1$ in Eqs.  \ref{eq:global_acoef} and \ref{eq:local_acoef}), these rate equations can be integrated to find the trajectory of the dynamical system. In this section, we explore this tractability to extract some insights from the model.

\subsection{Long-term strategy}
\label{sec:homix_lt}

Considering the dynamics of a single isolated subpopulation
, the dynamical equations for the average fractions $\rho_S$, $\rho_I$ and $\rho_R$ of susceptible, infectious and removed individuals can be written as:

\begin{align}
    \Dot{\rho_S} & = - \mu R_0 \, \rho_S \rho_I \cdot \, a(\rho_I, \rho_R) \label{eq:dyn_rho_s}\\
    \Dot{\rho_I} & = \mu R_0 \, \rho_S \rho_I \cdot \, a(\rho_I, \rho_R) - \mu \rho_I  \label{eq:dyn_rho_i}\\
    \Dot{\rho_R} & = \mu \rho_I, \label{eq:dyn_rho_r}
\end{align}

where $R_0 = \beta / \mu$ is the basic reproduction number and $\mu$ is now interpreted as the recovery rate. For the long-term strategy, $a(\rho_I, \rho_R) = (1 - (\rho_I + \rho_R))^k = \rho_S^k$, given that $\rho_S + \rho_I + \rho_R = 1$. As described in \cite{eksin2019systematic}, we can divide equation \ref{eq:dyn_rho_s} by equation \ref{eq:dyn_rho_r} to obtain a separable differential equation:

\begin{equation}
    \label{eq:dyn_parametric}
    \frac{\text{d}\rho_S}{\text{d}\rho_R} = - R_0 \rho_S^{k+1}
\end{equation}





For $k > 0$ and assuming that $\rho_R(0) = \rho_0 \geq 0$ and $\rho_I(0) = \delta \rightarrow 0$, its solution is given by:

\begin{equation}
    \rho_S = \left[ \frac{1}{(1 - \rho_0)^k} + k R_0 (\rho_R - \rho_0) \right]^{-\frac{1}{k}}
\end{equation}

Alternatively, writing $\rho_I$ as a function of $\rho_R$:

\begin{equation}
    \label{eq:sir_trajec_lt}
    \rho_I = 1 - \rho_R - \left[ \frac{1}{(1 - \rho_0)^k} + k R_0 (\rho_R - \rho_0)  \right]^{-\frac{1}{k}}
\end{equation}

\begin{figure}[h]
    \centering
    \includegraphics[width=0.49\textwidth]{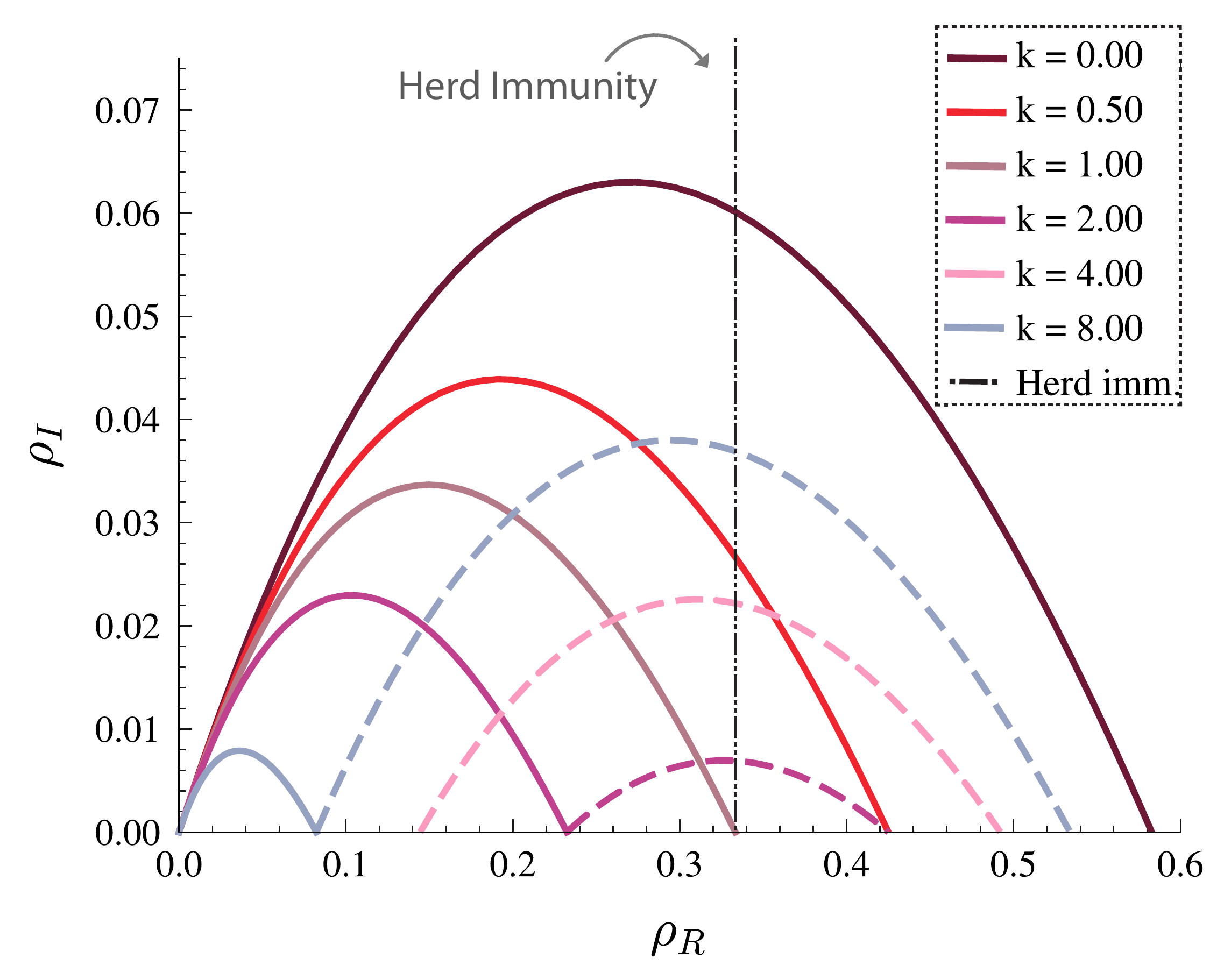}
    \caption{Analytical trajectories of the homogeneously mixed long-term (LT) strategy with system release after the first outbreak (i.e., complete lift of the contention measures). The trajectories are combinations of equations \ref{eq:sir_trajec_lt} (solid lines) and \ref{eq:sir_trajec_regular} (dashed lines) with $\rho_0$ set to the size of the first outbreak. $R_0$ is set to $1.5$.}
    \label{fig:homix_trajectories_release}
\end{figure}

Equation \ref{eq:sir_trajec_lt} represents the trajectory of the system in a $\rho_I$ \emph{vs} $\rho_R$ diagram, showing how the fraction of infectious individuals peaks as the prevalence increases in the LT strategy. Also, the final $\rho_R$ prevalence at the end of the outbreak can be found by solving equation \ref{eq:sir_trajec_lt} for $\rho_I = 0$ besides the trivial solution $\rho_R = \rho_0$. The solid lines in Figure \ref{fig:homix_trajectories_release} show such trajectories for different values of the response strength $k$, $\rho_0 = 0$ and $R_0 = 1.5$.

For $k = 0$, equation \ref{eq:dyn_parametric} solves as a simple SIR model which, for $\rho_R(0) = \rho_0$ and $\rho_I(0) = \delta \rightarrow 0$, leads to a $\rho_I$ \emph{vs} $\rho_R$ trajectory given by:

\begin{equation}
    \label{eq:sir_trajec_regular}
    \rho_I = 1 - \rho_R - (1 - \rho_0) e^{-R_0(\rho_R - \rho_0)}
\end{equation}

In the long-term strategy with $k>0$, the contact reduction coefficient $a(\rho_I, \rho_R)$ is a decreasing function of $\rho_R + \rho_I$, which in turn can only increase over time. This means that $a(\rho_I, \rho_R)$ will also always increase as if the disease contention measures are held forever. This is not reasonable to assume for the long term behavior of a real situation and, for this reason, we can consider that at some point after the main outbreak, $a(\rho_I, \rho_R)$ is set to $1$ again, which we call henceforth a \emph{system release} (representing a \emph{release} of the contention measures). 

For instance, consider that a system release occurs when the fraction of infectious individuals reaches a small value $\delta > 0$ after the main outbreak initiated. 
This comprises the fact that, during the time evolution given by equations \ref{eq:dyn_rho_s} to \ref{eq:dyn_rho_r}, $\rho_I(t)$ tends to but never actually reaches zero.
For the calculation of the $\rho_I$ \emph{vs} $\rho_R$ trajectories though, we can consider $\delta \rightarrow 0$. After the system is released, the trajectory follows Eq. \ref{eq:sir_trajec_regular} with $\rho_0$ set to the final size of the first outbreak. In Figure \ref{fig:homix_trajectories_release}, the dashed lines represent the secondary outbreaks produced after the system's release.

As observed in the figure, a system release can generate a second outbreak which, for higher values of $k$, can be greater than the first one. This happens because, although the long-term strategy effectively reduces the size of the first outbreak (both in $\rho_R$ and $\rho_I$), it may leave the system under its herd immunity threshold, and thus vulnerable to new outbreaks \cite{Lu2021Mar}. In this homogeneous model, the threshold for herd immunity is given by $\herdlim = 1 - R_0^{-1}$, meaning that for $\rho_R \geq \herdlim$ the fraction of infectious individuals can no longer increase. If the disease dies out before the prevalence reaches that value, the system will experience a new outbreak if the social distancing measures are relaxed and the disease is reintroduced in the population. Note that the critical value of $k$ for reaching herd immunity at the first outbreak is $k = 1$, for which the final outbreak size is exactly solved as $\rho_R^* = 1 - R_0^{-1}$, as noticed by Eksin and others \cite{eksin2019systematic}.








\subsection{System resets} 
\label{sec:homix_reset}

Besides \emph{releasing} the system after the first outbreak, we propose another way to work with multiple outbreaks in the LT strategy. What makes the $a(\rho_I, \rho_R)$ coefficient to be by default strictly decreasing is its dependence on $\rho_R$, which holds a ``long-term memory''. We can modify and possibly reset such memory by subtracting a constant $\rho_0$ from the prevalence that's considered for the contact reduction coefficient $a(\rho_I, \rho_R)$. If we do this right after the main outbreak and set $\rho_0$ to the recovered fraction at that time, then $a(\rho_I, \rho_R)$ is momentarily reset to 1 (i.e., no contention measures), but the system will still react in the event of another outbreak. We call this a \emph{system reset} (the memory of the population is reset, but the social distancing measures still apply), in contrast with the simpler \emph{system release} described in the previous section (in which the contention measures are completely removed). We are again assuming that $\rho_I$ is arbitrarily small at the end of an outbreak. 

During this new round of the dynamics, equation \ref{eq:dyn_rho_s} divided by equation \ref{eq:dyn_rho_r} yields the following differential equation:

\begin{equation}
    \label{eq:dyn_reset_parametric}
    \frac{\text{d}\rho_S}{\text{d}\rho_R} = - R_0 \cdot (\rho_S - \rho_0)^k \rho_S
\end{equation}

which is still separable, but for $\rho_0 > 0$ solves into a less insightful expression:

\begin{equation}
    \label{eq:mem_reset_trajec}
    R_0 \cdot (\rho_R - \rho_0) = P_{\rho_0, k}(\rho_S)
\end{equation}

where we define:

\begin{align}
    \label{eq:p_rhos}
    P_{\rho_0, k}(\rho_S) & = \int_{\rho_S}^{1 - \rho_0} \frac{\text{d}u}{u (u + \rho_0)^k} = \\ 
    & = \left. - \frac{1}{k(u+\rho_0)^k} \left(  1 + \frac{\rho_0}{u} \right)^k {}_2{F_1} \left(k, k; k+1, -\frac{\rho_0}{u} \right)   \right| _{\rho_S}^{1 - \rho_0}  \nonumber
\end{align}

where ${}_2F_1$ is the Gaussian hypergeometric function. Equations \ref{eq:mem_reset_trajec} and \ref{eq:p_rhos} provide $\rho_R$ for any given $\rho_S$. For each of these values, $\rho_I$ is calculated as $\rho_I = 1 - \rho_S - \rho_R$, which finally allows the construction of a $\rho_I$ \emph{vs} $\rho_R$ trajectory.

\begin{figure}[h]
    \centering
    \includegraphics[width=0.49\textwidth]{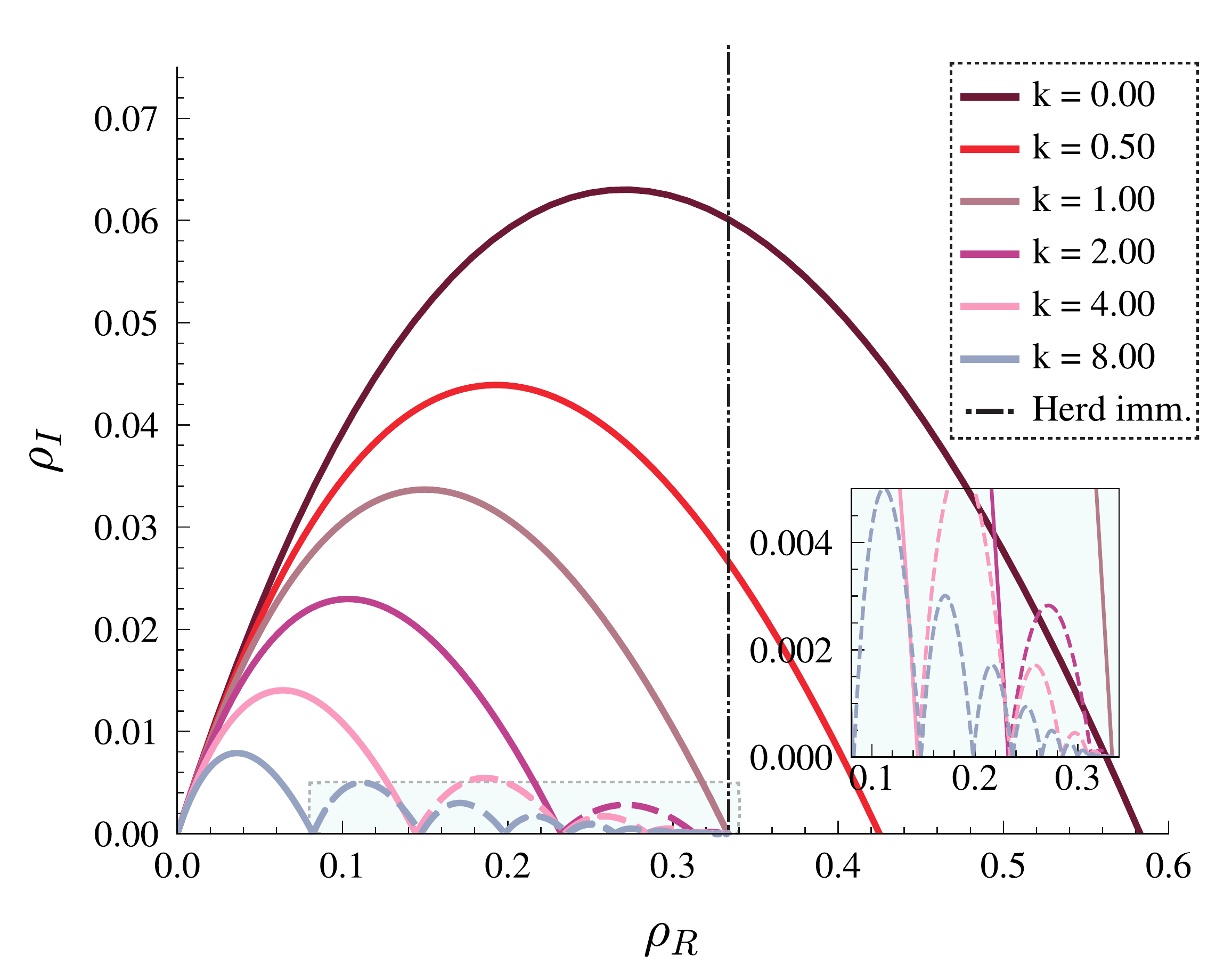}
    \caption{Analytical trajectories of the homogeneously mixed long-term (LT) strategy with system resets. The first outbreak (solid lines) follows equation \ref{eq:sir_trajec_lt}, while subsequent outbreaks (dashed lines) are traced using equations \ref{eq:mem_reset_trajec} and \ref{eq:p_rhos}. 
    The inset shows a sub-region of the main plot (indicated by a blue dotted rectangle) where secondary outbreaks are more easily visible, showing how there may be multiple and successively smaller peaks.
    $R_0$ is set to 1.5.}
    \label{fig:homix_trajectories_reset}
\end{figure}

Figure \ref{fig:homix_trajectories_reset} shows the trajectories of the model with system resets, for different values of $k$. The first outbreak of each execution is represented by solid lines, and are traced using equation \ref{eq:sir_trajec_lt}. For the executions that did not achieve the herd immunity threshold (which is shown as a black dot-dashed line), subsequent outbreaks are represented by dashed lines and traced using equations \ref{eq:mem_reset_trajec} and \ref{eq:p_rhos}. In this scenario, stricter contention measures (that is, higher $k$) always cause smaller infectious peaks, but can generate more secondary outbreaks, making it more difficult to comply with the measures. 
The inset of Figure \ref{fig:homix_trajectories_reset} shows in more detail the secondary outbreaks for $k = 2, 4$ and $8$, where we can see that there are multiple outbreaks with decreasing peak size. 

\subsection{Short-term strategy}

For the short-term (ST) strategy, the coefficient of contact reduction depends only on $\rho_I$, being written as $a(\rho_I) = (1 - \rho_I)^k = (\rho_R + \rho_S)^k$. This breaks the separability of the differential equation obtained by dividing equations \ref{eq:dyn_rho_s} and \ref{eq:dyn_rho_r}, leaving no simple method to solve it for arbitrary values of $k$. We can still use a classic Runge-Kutta of order 4(5) \cite{dormand1980family} to integrate the equations in time, and plot the $\rho_I$ \emph{vs} $\rho_R$ trajectories.

\begin{figure}[h]
    \centering
    \includegraphics[width=0.49\textwidth]{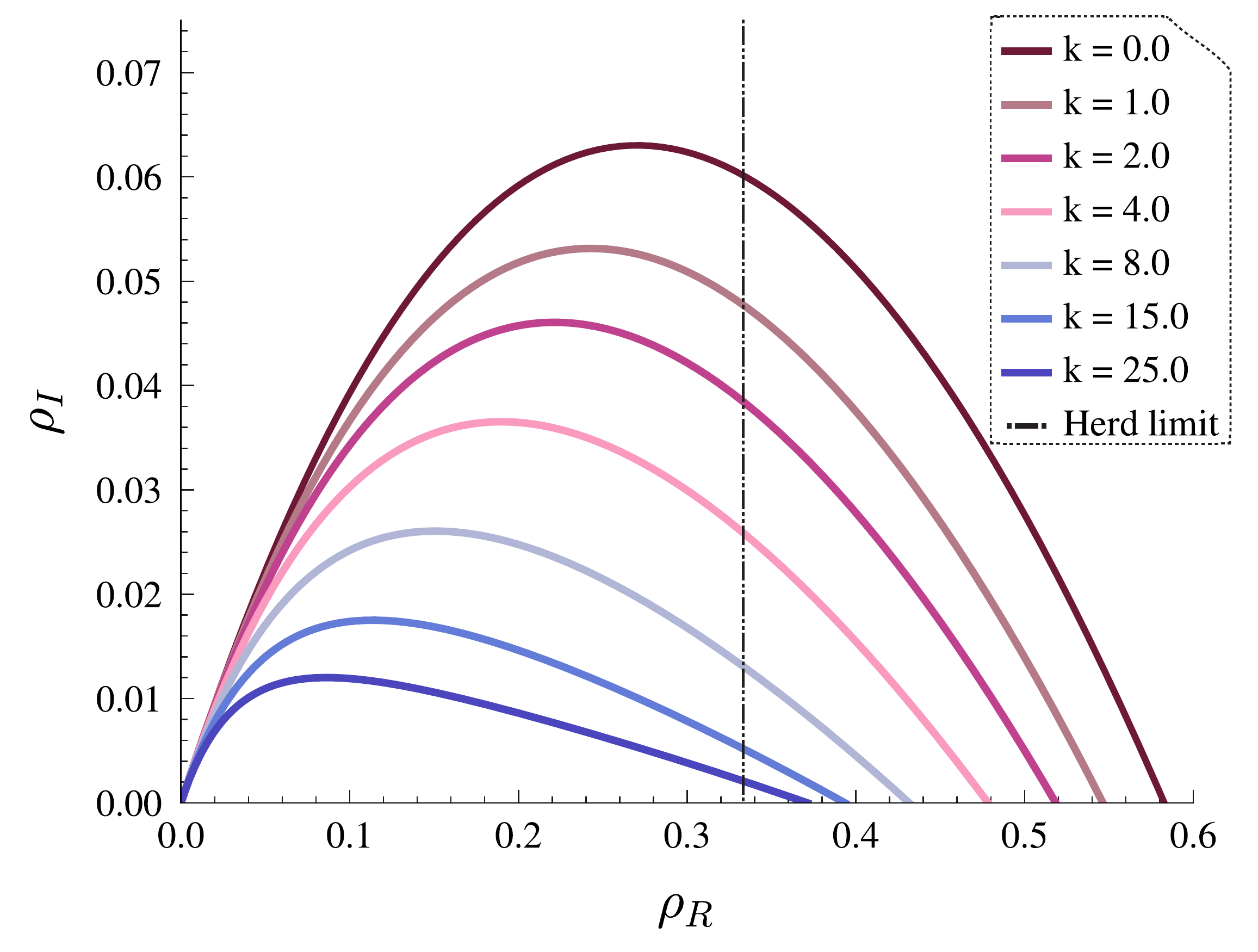}
    \caption{Numerical trajectories of the homogeneously mixed short-term (ST) strategy, obtained from Runge-Kutta integration of equations \ref{eq:dyn_rho_s} to \ref{eq:dyn_rho_r} starting with $\rho_I(0) = 1\cdot 10^{-5}$ and $\rho_R(0) = 0$. $R_0$ is set to 1.5. Unlike the LT strategy, the system always reaches the herd immunity condition in a single outbreak, without considering any modification to the baseline model.}
    \label{fig:homix_trajectories_st}
\end{figure}

Figure \ref{fig:homix_trajectories_st} shows the numerically solved trajectories of the system for different values of $k$. The main difference with respect to the LT strategy is that the system eventually reaches the herd immunity condition within the first and only outbreak. Higher values of $k$ reduce the size of the peak in $\rho_I$, at the expense of larger time-span in which the contention measures have to be sustained. This cannot be visualized in Figure \ref{fig:homix_trajectories_st} as the time parameter is implicit, but we further explore the interplay between peak size and time span in section \ref{sec:loc-vs-glob_st}. Note also that even for very large values of $k$ the final fraction of removed individuals is beyond the herd immunity threshold.


\section{Global strategies and the heterogeneity of local features}
\label{sec:heterogeneities}

The previous analysis has shown that, even if local extinction might be achievable with social distancing, it renders the system vulnerable to further reintroductions of the pathogen. Thus, if mobility between subpopulations is allowed, there could be spill overs from those with still ongoing outbreaks to those which contained the epidemic. To study this type of events, we now focus on the proposed model in a metapopulation using Monte Carlo simulations, with the algorithm described in section \ref{sec:description}. For the metapopulation, we use a random geometric network (RGN) of $\nnodes = 50$ subpopulations, constructed in a square space of length $1$ and connecting subpopulations that are closer than $d = 0.25$. At this initial point, all links are unweighted and reciprocal. This gives an expected average degree of $\nnodes \pi d^2 \approx 9.82$, though the specific realization we used through this paper has an average degree of $7.2$. We choose such a network topology because it emulates the spatial distribution of cities in small to mid scale, where size hierarchy and long-range links are not very present.

Once this unweighted graph is constructed, we set the initial population of each subpopulation $i$ proportional to its degree $k_i$, according to $N_i(0) = \lfloor \npop k_i / (2M) \rfloor$, where we set $\npop{} = 10^7$. This makes the overall population to be not exactly but very close to $\npop{}$, only deviated due to truncation. Then we set the weights of existing links between subpopulations $i$ and $j$ as $T_{ij} = N_i(0) N_j(0) / \npop = T_{ji}$. Within this scheme, the local population sizes fluctuate around $N(0)$ over time, as explained before. Moreover, the most connected subpopulations are also most populous ones, though the RGN is reasonably homogeneous in this sense. For consistency of the results, we also use a fixed subpopulation as the seed of the disease, seeding 10 infectious individuals at the beginning of the simulations, with all other subpopulations starting with susceptibles only. The seeded subpopulation was chosen to be around the center of the square space.

\subsection{Long-term (LT) global strategy}

\begin{figure}[h]
    \centering
    \includegraphics[width=0.48\textwidth, trim=0 0 0 0, clip]{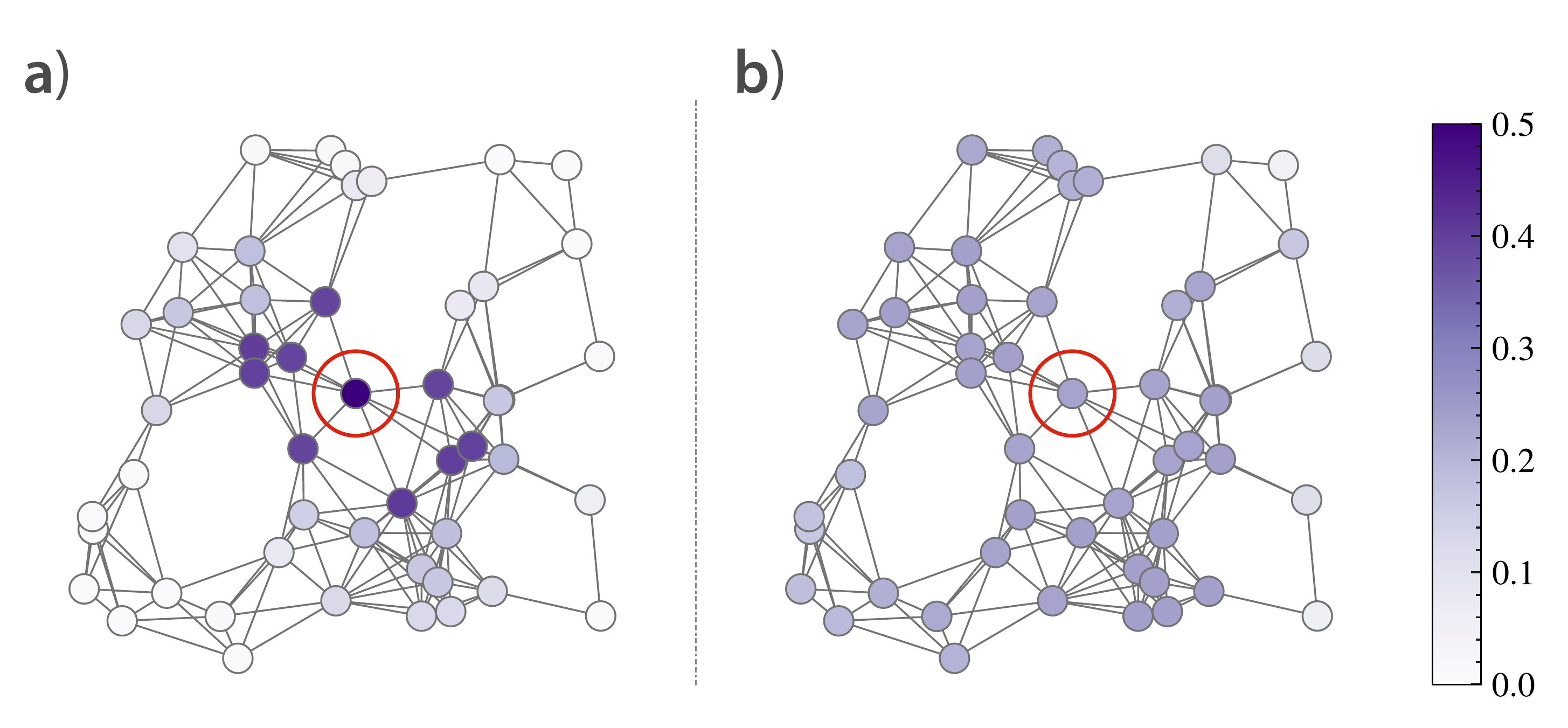}
    \caption{Comparison of the average outbreak size in each subpopulation between a) global LT strategy and b) global constant response with threshold. For the LT strategy, a response strength $k = 2$ was used, while the constant response uses $a_0 = 0.78$ and a threshold of $\ithreshold{} = 10^{-3}$, meaning that the contact reduction is triggered after $10^4$ overall infections (out of $\npop{} = 10^7$ individuals) are registered. Both simulations use $\tau = 1\cdot10^{-3}$ as the travel parameter.}
    \label{fig:netw_plot_lt_rmax}
\end{figure}

\begin{figure}[h]
    \centering
    \includegraphics[width=0.48\textwidth, trim=0 0 1100 30, clip]{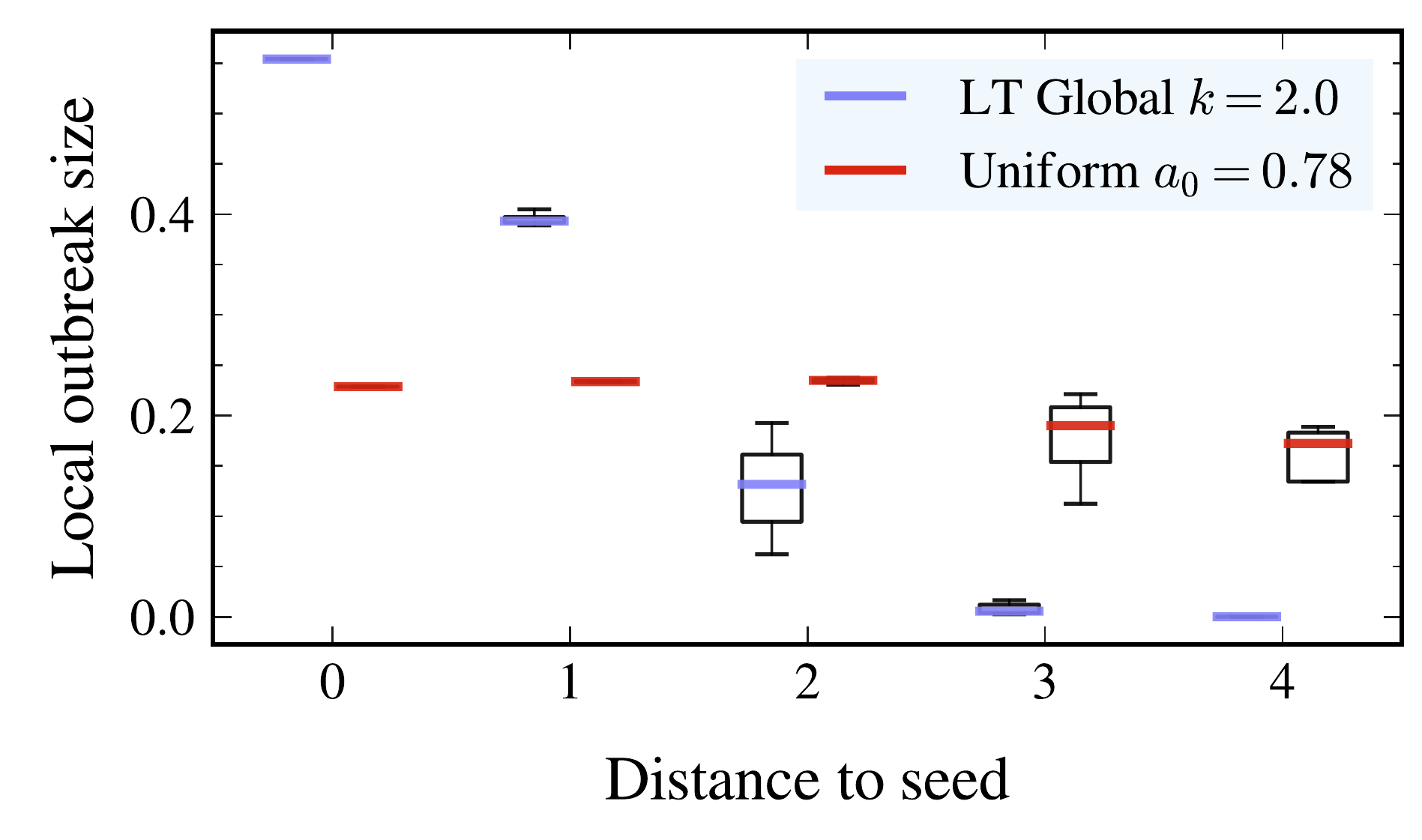}  
    \caption{Local outbreak size (given by the maximum of $R_i / N_i$ over time) in each subpopulation as a function of its shortest path length to the seed, for global LT strategy with response strength $k = 2$ (circles) and for constant response with factor $a_0 = 0.78$ (squares). The y-axis values represent averages over $1000$ independent executions. The boxes show the inner quartiles of each set. The mobility coefficient is set to $\tau = 1\cdot10^{-3}$.}
    \label{fig:subplots_lt_rmax}
\end{figure}

For the long-term (LT) strategy with global information, we can compare the outcomes of the simulations with those of a constant response after a threshold (explained in section \ref{sec:description_epid}). Figure \ref{fig:netw_plot_lt_rmax} shows the outbreak size (total number of infected individuals) in each subpopulation for the two strategies, each one averaged over $1000$ independent executions. The constant coefficient of contact reduction $a_0 = 0.78$ was chosen to yield a global outbreak size of $\sim0.20$, approximately equal to that obtained with the LT global strategy with $k = 2$. We notice that the LT strategy provides more heterogeneous outcomes between subpopulations compared to the constant response. Specially, as it can be seen in Figure \ref{fig:subplots_lt_rmax}, subpopulations that are farther from the seed tend to have smaller outbreak sizes. This happens because, in the global LT strategy, the intensity of contention measures due to public awareness is the same for all subpopulations and increases with time, thus subpopulations that are seeded later have smaller effective reproduction numbers. 


The fact that our proposed model for behavioral responses with global strategy produces more heterogeneous outbreak sizes is notorious, as this is generally observed after outbreaks of real diseases. In particular, this situation is compatible with the one observed in 2020 during the COVID-19 pandemic in those countries that imposed strict global lockdowns even if only part of the country was severely affected \cite{Pollan2020Aug,Riccardo2020Dec,Chinazzi2020Apr}. In contrast, a uniform strategy would lead to more homogeneous outcomes.

\FloatBarrier
\section{The efficiency of local and global strategies}

We now describe a framework to compare local and global strategies in terms of their efficiency, characterized by the costs and benefits of each strategy. Direct comparison of simulations with global and local strategies using the same response strength $k$ is inappropriate, as global strategies require greater values of $k$ to yield similar effects. We therefore define two metrics, one for the cost and another for the effectiveness, and compare local and global strategies in parametric plots of such metrics (with $k$ as an implicit parameter). As long-term and short-term strategies are qualitatively different, we apply different metrics to characterize each one.

\subsection{Short-term (ST) strategy}
\label{sec:loc-vs-glob_st}

The short-term strategy, in which the contact reduction coefficient only responds to the (either local or global) density of infectious individuals, is characterized by a slow progression of the system towards its herd immunity.
A higher value of the response strength $k$ represents a more effective response, which reflects into a smaller prevalence peak (i.e., the maximum $\rho_I$) yet longer outbreak duration. The outbreak size (i.e., maximum of $\rho_R$) however does not vary much with $k$, as it essentially depends on the herd immunity limit of the system. Therefore, the infectious peak size $\peaksize$ is a reasonable measure of effectiveness in this case, with smaller peaks attributed to more effective strategies.

Higher values of $k$ also imply more time in which contention measures are active to control disease propagation. This translates into $\acoef$ being smaller than $1$ during longer times. We can quantify ``how much and for how long'' the contention measures are applied with the quantity:

\begin{equation}
    \label{eq:aimpulse}
    \aimpulse{}_i = \sum_{t = 0}^\infty (1 - \acoef(t))
\end{equation}

which accounts for the intensity and time span of the contact reduction. This provides, in a broad sense, a metric for the cost of the short term strategy, considering that contention measures typically bring costs to the population. Notice that these metrics must be calculated for each subpopulation.

\begin{figure}[h]
    \centering
    \includegraphics[width=0.47\textwidth]{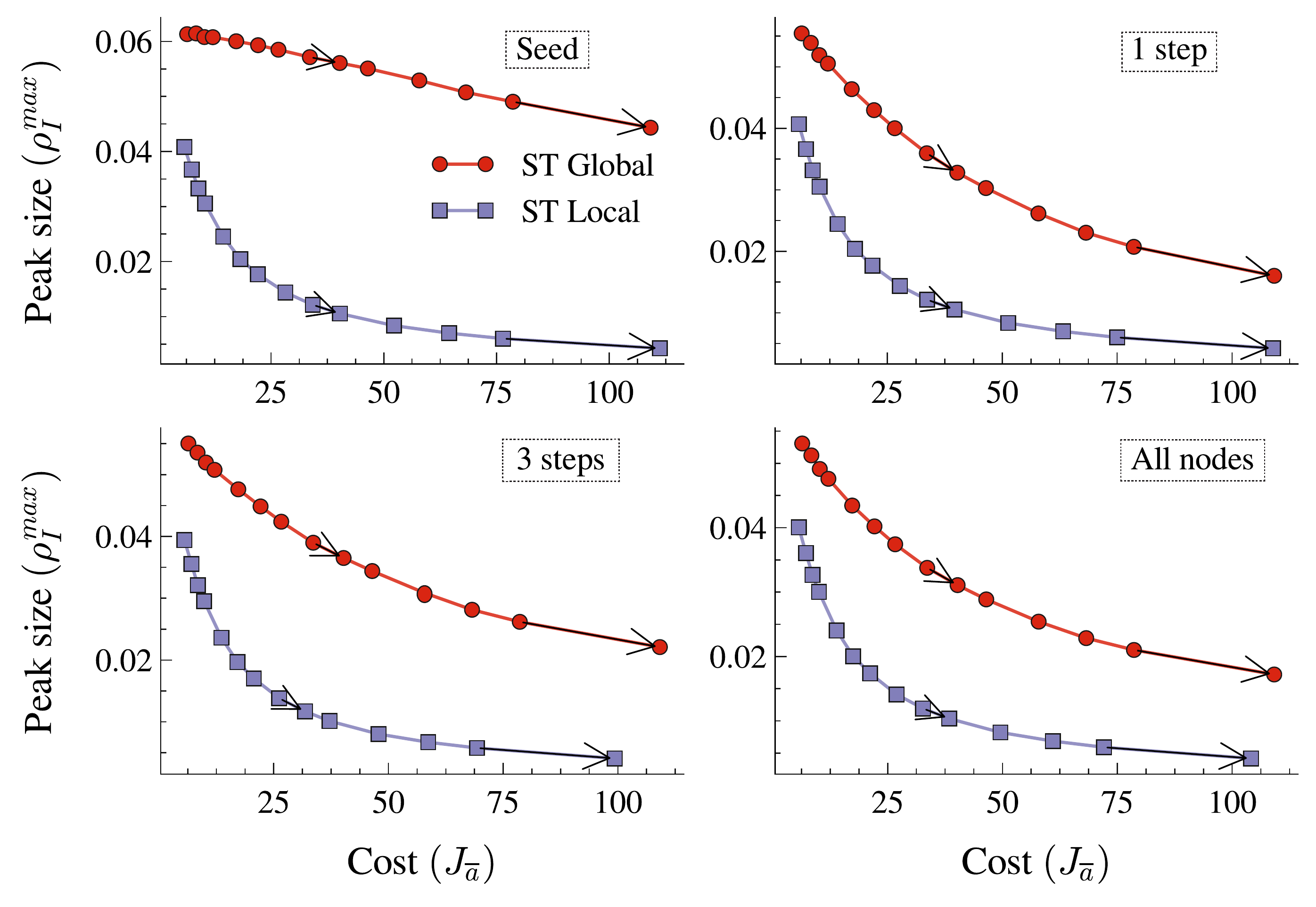}
    \caption{Parametric curves for the effectiveness and cost of local and global ST strategy, for values of $k$ ranging from $3$ to $90$. The first plot is for the seeded subpopulation, whereas other plots are arithmetic averages over sets of subpopulations according to their shortest path length to the seeded subpopulation. The last plot is an arithmetic average over all subpopulations. The arrows indicate the direction towards which $k$ increases. The mobility factor is set to $\tau = 1\cdot 10^{-3}$. $\peaksize$ is the maximum over time of the local infection prevalence $\rho_I$, while $\aimpulse$ is given by equation \ref{eq:aimpulse}.}
    \label{fig:feats_st_tau1e-3}
\end{figure}

For the same population and RGN structure described in section \ref{sec:heterogeneities}, we compare local and global strategies with respect to $\peaksize$ and $\aimpulse$ for simulations with different values of $k$. Figure \ref{fig:feats_st_tau1e-3} shows the plots of these metrics against each other, grouped by sets of subpopulations according to their distance to the seeded subpopulation. The metric values are averaged over $1000$ independent executions.

From the plots, it is clear that local strategies outperform global ones for all subpopulations, as the former produces much smaller peaks sizes for similar values of $\aimpulse$ in the considered range. This means that, for the ST strategy, the use of local information about contagions 
is more efficient than using a unified, global strategy. This happens because using global information may not be well suited for the epidemic situation of a given subpopulation. For instance, subpopulations that are far from the seed effectively apply the contention measures before the epidemic arrives, but as these measures are relaxed at some moment (when the global prevalence decreases), these subpopulations will still undergo an intense outbreak, given that they were only subject to minor outbreaks and the majority of the individuals in these subpopulations are susceptible.

Although we only display the results for a single value of the mobility coefficient $\tau = 10^{-3}$, the same conclusions are obtained for $\tau = 10^{-2}$ and $10^{-4}$, but the advantage of local strategies over global ones is more pronounced with lower mobility rates. This is expected, because with lower $\tau$ values, the dynamics of each subpopulation is less coupled by mobility, making local strategies more practical.


\subsection{Long-term (LT) strategy}

In the long-term strategy, the overall intensity of contention measures increases with time (though it can locally
decrease due to migration), as they are proportional to both I and R densities. As shown in section \ref{sec:analyt} for homogeneous mixing, for sufficiently high $k$, the epidemic spreading is halted before herd immunity is achieved, leaving the system vulnerable to secondary waves in case of a system release without other immunization policies. We can still characterize the effectiveness of the strategy in this first wave, either by its outbreak size ($\outbsize$) or by the peak size ($\peaksize$). For consistency with section \ref{sec:loc-vs-glob_st}, we chose to work with $\peaksize$ as well. 

For the cost of the strategy, $\aimpulse$ (given by equation \ref{eq:aimpulse}) does not provide a reliable measure, as in this case $\acoef$ does not approach $1$ again at the end of the main outbreak, making $\aimpulse$ overly sensitive to the time taken until the epidemic vanishes. We choose instead to simply work with $\abarmax = 1 - \min{a(t)}$, which is the maximum level of contact reduction adopted by the subpopulation during the simulation. This metric disregards the temporal evolution of $\acoef$, but still provides a number that is proportional to the level of contention measures adopted by the subpopulations without the sensibility issue of $\aimpulse{}$.

\begin{figure*}
    \centering
    \includegraphics[width=0.70\textwidth]{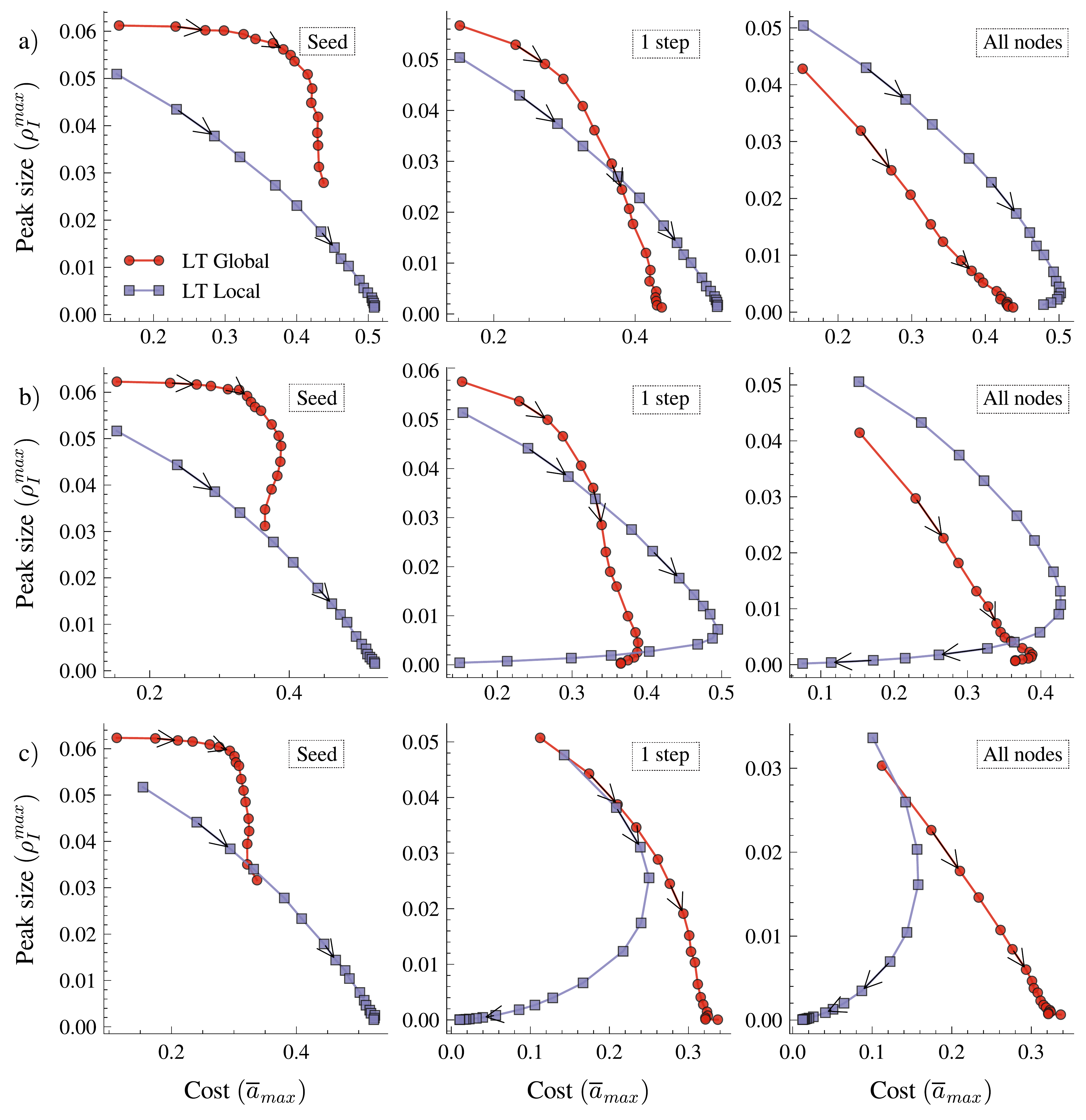}
    \caption{Parametric curves for the effectiveness and cost of local and global LT strategy, for the mobility parameter (a) $\tau =10^{-2}$, (b) $\tau =10^{-3}$ and (c) $\tau =10^{-4}$. The $k$ values range from $0.25$ to $50$. The leftmost panels represent the seeded subpopulation, the middle panels represent an arithmetic average over its immediate neighbors, and the rightmost panels show the averages over all subpopulations. The arrows indicate the direction towards which $k$ increases. The cost $\abarmax$ is the maximum value of $1 - \acoef{}$ reached during the simulation.}
    \label{fig:feats_lt}
\end{figure*}

Using the same RGN setup, we find that the cost \emph{vs} effectiveness curves are more complex for the LT strategies than for the ST ones. Figure \ref{fig:feats_lt} shows the $\outbsize$ x $\abarmax$ parametric curves for local and global strategies, and for three values of the mobility master parameter $\tau = 10^{-2}$, $10^{-3}$ and $10^{-4}$. In this case, there is not a clear advantage of local strategies over global ones, and this depends both on mobility levels and the distance to the seed. For instance, when $\tau = 10^{-2}$ (Figure \ref{fig:feats_lt}.a)), a local strategy is better for the seed, while it is generally worse for more distant subpopulations. For immediate neighbors of the seed (one step away), the curves cross each other, making the optimal strategy to depend on the value of $k$.

For lower mobility values (Figure \ref{fig:feats_lt}.b) and c)), another interesting feature is evident: the curves for the local strategies are not monotonic with the cost, meaning that two strategies (given by two different values of $k$) may have the same cost but notably different effectiveness. This happens because the LT strategy can affect the invasion threshold and attack rate of the system, preventing some subpopulations from being reached by the disease for sufficiently high $k$. In this case, as the strategy is based on local prevalences, these subpopulations will not have to implement contention measures, which explains the decrease in the strategy cost.

\section{Secondary outbreaks in the global long-term strategy}
\label{sec:reset_strategy}

We adapt the metapopulation model with a long-term global strategy to consider system resets, as explained in section \ref{sec:homix_reset}. This allows us to analyze secondary outbreaks caused by a relief in the contention measures, represented here as a reset in the ``memory'' of the contact reduction mechanism (that is, the value of $\rho_R$ in the $\acoef$ coefficient).

The system resets are implemented as follows: 
the contention measures are activated when the overall fraction of infectious individuals $\rho_I$ surpasses an activation threshold $\ithreshold^{\text{in}} = 10^{-4}$ (activation threshold), and then deactivated (possibly after an outbreak) if $\rho_I$ goes under another threshold $\ithreshold^{\text{out}} = 0.8\cdot10^{-4} < \ithreshold^\in$. At each activation event, the ``memory'' of the system is reset, that is, the coefficient of contact reduction is given by $\acoef{}(t) = (1 - (\rho_I(t) + \rho_R(t) - \rho_0))^k$, with $\rho_0$ set to the value of $\rho_R$ at the moment of the activation event. This mechanism is slightly different from that used in section \ref{sec:homix_reset} for the rate equations model, but the distinction $\ithreshold^{\text{out}} < \ithreshold^{\text{in}}$ is needed in stochastic simulations so that small fluctuations around $\ithreshold$ do not increase the number of outbreaks.
To control the number of secondary outbreaks, we also limit the number of system reset events to $10$. If this number of reset events is reached before the extinction of the epidemics, the system is then released, that is, $\acoef{}$ is permanently set to $1$. 



\begin{figure}[h]
    \centering
    \includegraphics[width=0.48\textwidth, trim=0 0 0 0, clip]{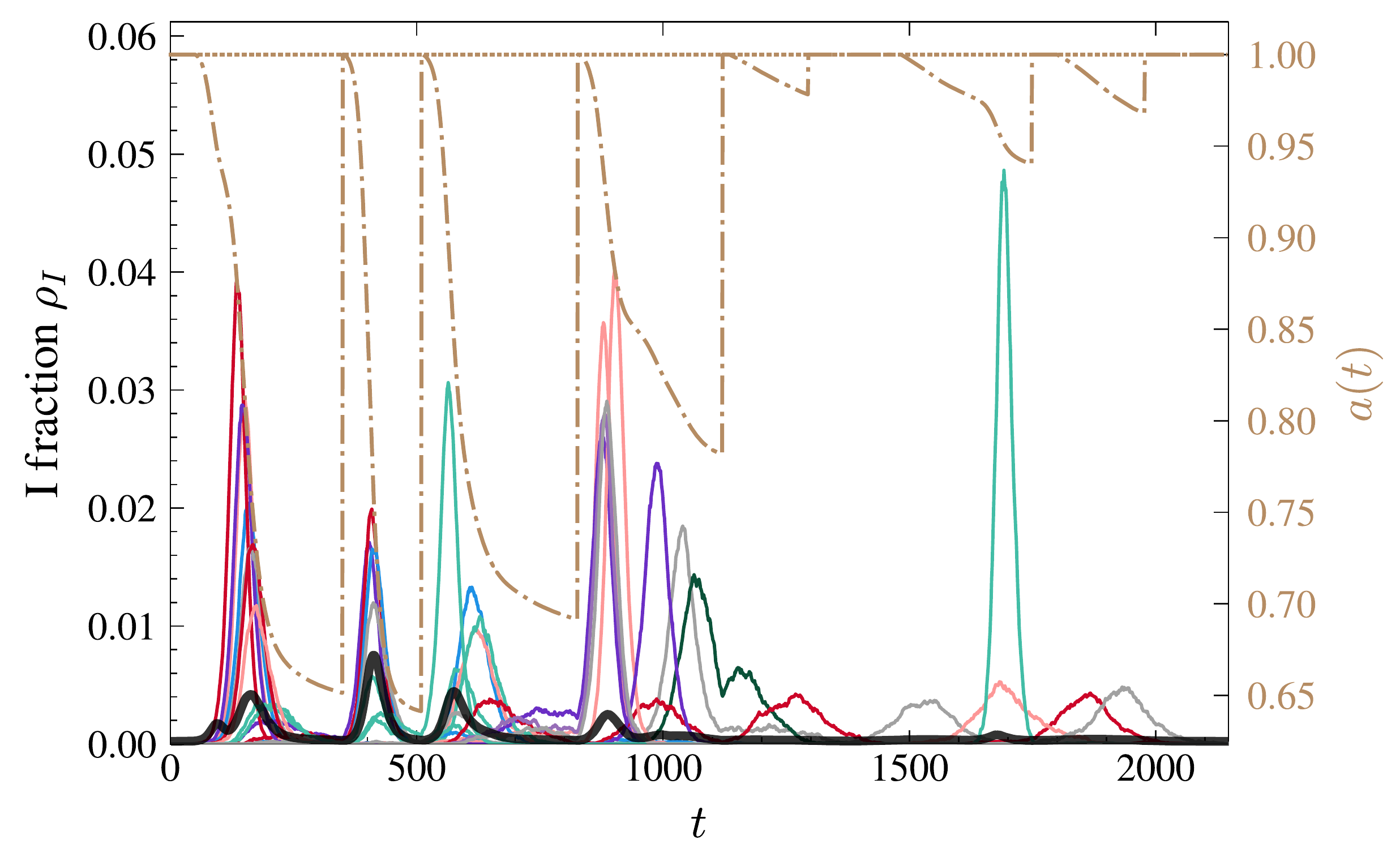}
    \caption{Time series of the fraction of infectious individuals in each subpopulation (colored lines), as well as in the whole population (black line). The global value of $a(t)$ is also shown as a red dot-dashed line (with the upper limit $a(t) = 1$ as a red thin horizontal line). The parameters (described in text) are set to: $k = 5$, 
    $\ithreshold^\text{in} = 10^{-4}$, $\ithreshold^\text{out} = 0.8\cdot10^{-4}$, $\tau = 10^{-3}$.  
    }
    \label{fig:reset_tseries}
\end{figure}

\begin{figure*}[t]
    \centering
    \includegraphics[width=0.95\textwidth]{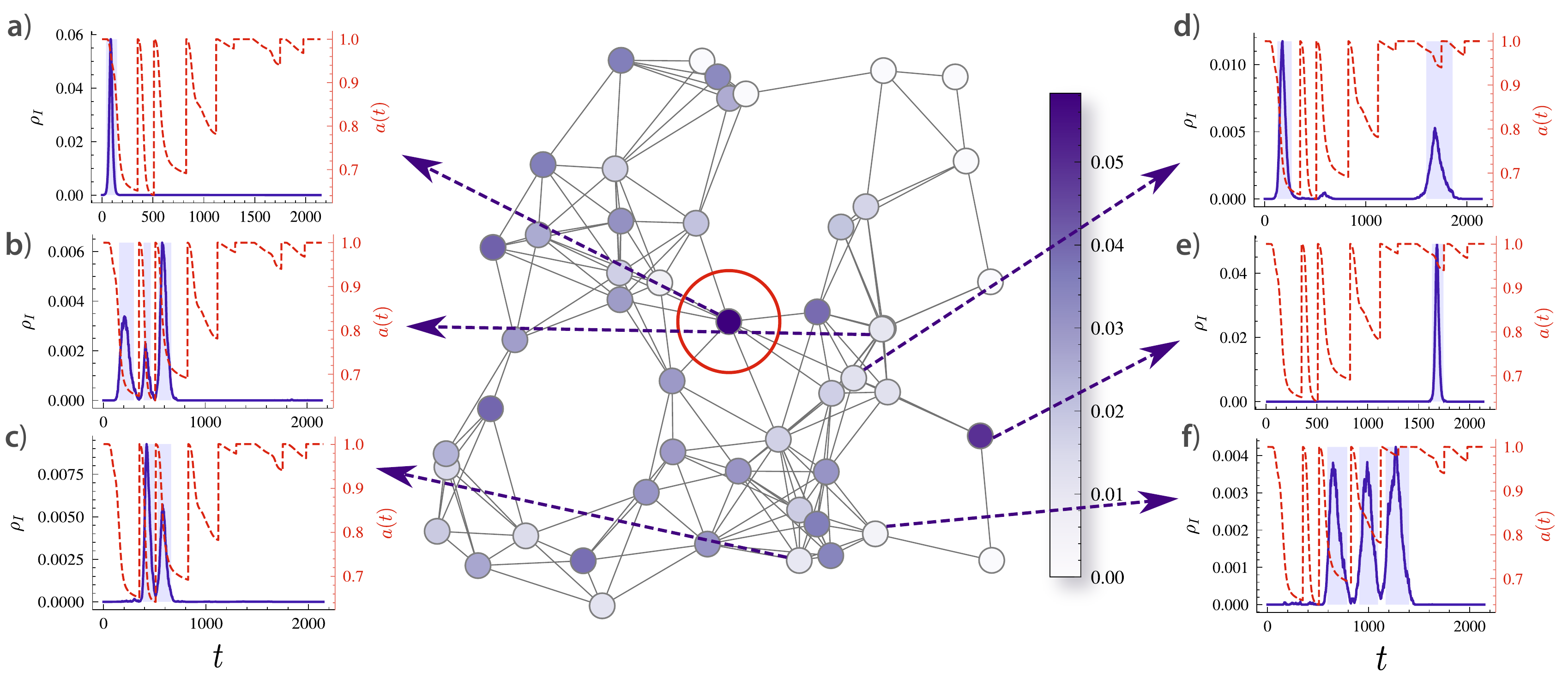}
    \caption{ Map of the final outbreak sizes ($\outbsize$) of each subpopulation, for a single execution with the memory reset mechanism (main panel). The panels show the time series of $I_i / N_i$ for some sample subpopulations, showing that the mechanism introduces diversity in the epidemic trajectory over the population. The average number of detected outbreaks for this run was $1.20$.}
    \label{fig:reset_node_tseries}
\end{figure*}

\begin{figure}
    \centering
    \includegraphics[width=0.48\textwidth]{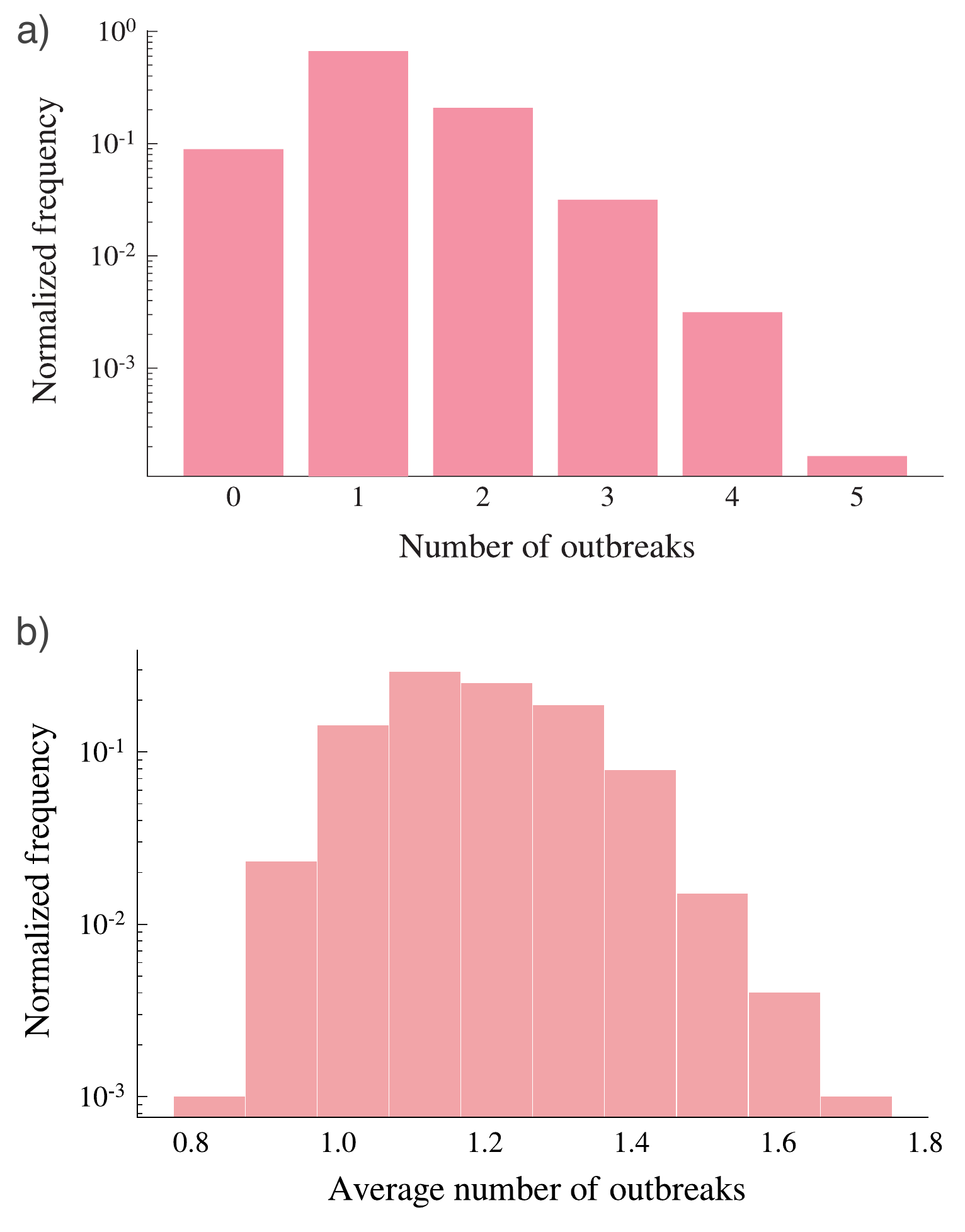}
    \caption{Histograms of (a) the number of local outbreaks (at subpopulation level) for all subpopulations, and (b) of the average number of outbreaks in each execution (the average is taken over subpopulations, but not over executions). The parameters of the simulations are set to: $k=5$, 
    $\ithreshold^\text{in} = 1\cdot10^{-4}$ and $\ithreshold^\text{out} = 0.8\cdot10^{-4}$. The number of system reset events was constrained to at most 10.
    }
    \label{fig:reset_num-outbreaks}
\end{figure}

In Figure \ref{fig:reset_tseries}, we show the time series of the fraction of infectious individuals of each subpopulation (colored thin curves), as well as the global fraction of infectious individuals (black shaded curve) and the value of $\acoef{}$ over time (red dashed curve) for a typical execution of the model. Using the same RGN population as in previous sections, we set the response strength as $k = 5$ and the mobility coefficient to $\tau = 10^{-3}$. Each vertical ascent of the $\acoef{}$ value represents the occurrence of a system reset. We notice that, at each reset, a new set of local outbreaks occurs, raising again the overall incidence and leading the system to readopt contention measures. Differently from the homogeneously mixed case (as in Figure \ref{fig:homix_trajectories_reset}), the secondary outbreaks can be greater in peak size than the first one. This feature is an essential difference between homogeneously mixed populations and structured ones like a metapopulation, and occurs because some (usually many) subpopulations are not reached by the primary outbreak.

We can better understand this feature by looking at the local incidence. Figure \ref{fig:reset_node_tseries} shows a network map of the greatest peak size (given by $\peaksize{}$) of each subpopulation, along with the time series of the fraction of infectious individuals for some selected subpopulations (lateral panels). The curve of panel (a) corresponds to the seeded subpopulation, displaying a single and pronounced peak in the early stage of the process. Other subpopulations may also present a single main outbreak (as in panel (e)), two (separate or close) outbreaks ((d) and (c)) or even more outbreaks ((b) and (f)). This spatial heterogeneity driven by contention measures has been observed in several locations during the COVID-19 pandemic \cite{Costa2020Dec,Sun2020Dec,Dong2020May,Starnini2020Jun}.

The features of a simulation with system resets are better presented through a single, typical execution of the simulation. We show that the observed pattern of multiple outbreaks is a solid feature of our model by counting the number of outbreaks of each time series, then plotting histograms for $1000$ independent executions. We use a simple algorithm to determine the number of outbreaks of each local time series, described as follows: every time the fraction of infectious individuals crosses a given threshold $\outbthresin = 1\cdot10^{-3}$ from bellow and, posteriorly, another threshold $\outbthresout = 0.5 \outbthresin < \outbthresin$ from above, an outbreak is accounted, and the time interval between these two crossings is regarded as a single outbreak. The difference between the two thresholds reduces spurious detection of outbreaks due to stochastic fluctuations, though this may still marginally occur. 

Figure \ref{fig:reset_num-outbreaks} shows the statistics of accounted local outbreaks. On panel (a), the number of outbreaks for each subpopulation and each of the $1000$ executions is put into a single histogram, showing that most subpopulations present a single outbreak, but often higher number of outbreaks also occur. On panel (b), we take the average number of outbreaks over all subpopulations in a single execution, then plot a histogram for the executions. From it we see that the typical average number of outbreaks is between $1$ and $1.4$, meaning that most executions present at least some subpopulations that undergo multiple outbreaks.

\section{Conclusions}

We have presented a model that incorporates dynamical behavioral responses to epidemics, which could be driven by governmental policies or by the endogenous response of individuals (e.g. fear of infection), in the context of metapopulations. The emerging features of the model are very rich, showing a complex landscape of outcomes depending on the implemented strategy.

First, we have shown that for isolated populations, strong social distancing measures (LT) are able to ablate the epidemic, but render the system vulnerable. Indeed, as long as the prevalence is below the herd immunity threshold, a reintroduction of the pathogen after the measures are lifted will inevitably lead to a second wave. On the other hand, soft social distancing measures (ST) contain the exponential growth of the epidemic and leave the system in a state in which the prevalence is over the herd immunity threshold, preventing further outbreaks. Note, however, that for a severe disease in which the infection fatality rate is substantial, larger prevalence implies a larger number of deaths. Thus, stronger social distancing policies will reduce the number of deaths at the expense of leaving the system vulnerable, while softer measures will control the spread but increase the number of deaths. These observations are particularly important in the context of metapopulations. Since subpopulations are not isolated, in the absence of additional control measures, the disease might be seeded again in disease-free areas by individuals who where infected in other regions. Thus, besides long and short term strategies, we also need to incorporate whether the contention measures use global or local information.

We compared the LT and ST strategies of our model, in an attempt to address the question of whether the contention measures are more efficient if implemented and managed locally (at each subpopulation) or globally (equally to the whole population). For the ST strategy, we measured its cost by the  $\aimpulse$ metric that quantifies the intensity and duration of the contention measures, and used the maximum number of simultaneous infections (i.e., the peak size) to quantify the strategy's performance. We showed that the local strategy always outperformed the global one, which is valid for any value of the mobility parameter. For the LT strategy, we quantified the cost by the maximum intensity reached by the contention measures, and the performance by the peak size. In this case, the cost/benefit relationship was more complex, depending on the mobility rate and the distance to the seeded subpopulation. Global strategies are generally better for distant subpopulations (with respect to the seed).  This situation was precisely the one that emerged in the first wave of infections in 2020 during the COVID-19 pandemic in countries that imposed country-wide lockdowns even if only one region was severely affected \cite{Pollan2020Aug,Riccardo2020Dec,Chinazzi2020Apr}. Yet, this is reversed if the mobility is low enough so that the local strategy can stop the epidemics before these nodes are reached. For the seeded subpopulation and its neighbors, local strategies are typically preferred. For this type of strategy, therefore, the choice between global and local strategies is not trivial and should be addressed appropriately. 

However, it is important to notice that long-term strategies assume a permanent adoption of contention measures, which leaves the system vulnerable to secondary outbreaks if these measures are lifted. We address this by implementing memory resets which, after outbreaks, instantly lifts the intensity of contention measures but leaves its mechanism active. Due to the residual presence of infectious individuals, the system faces secondary outbreaks. Thus, subpopulations that experienced a milder first wave, might have worse outcomes in subsequent waves. Besides, regardless of the contention measures, once they are lifted the system keeps progressing towards the herd immunity threshold, but each population might experience waves of different intensity at different times. The spatial heterogeneity driven by contention measures depicted in this paper is compatible with the one observed during the COVID-19 pandemic in several regions of the world \cite{Costa2020Dec,Sun2020Dec,Dong2020May,Starnini2020Jun}. 

In summary, we have shown than even in a simple metapopulation model a very complex scenario emerges. Depending on the mobility and distance to the seed, local vs global strategies yield different results and introduce different heterogeneities. As such, the full complexity of human gatherings and behaviors should be accounted for to effectively deal with emerging diseases.


\section{Acknowledgments}

P.C.V. acknowledges the financial support of FAPESP through grants 2016/24555-0 and 2019/11183-5. Y.M. acknowledges partial support from the Government of Aragón, Spain through grant E36-20R, by MINECO and FEDER funds (grant FIS2017-87519-P). A.A. and Y.M. acknowledge support from Intesa Sanpaolo Innovation Center.  F.A.R. acknowledges CNPq (grant 309266/2019-0) and FAPESP (grant 19/23293-0) for the financial support given for this research.

\FloatBarrier
\bibliographystyle{unsrt}
\bibliography{bibliography.bib}

\end{document}